\documentclass[final,times]{elsarticle}
\usepackage{graphicx}

\usepackage{amssymb}

\begin{document}

\begin{frontmatter}


\author{V.Celebonovic\corref{cor1}}
\ead{vladan@ipb.ac.rs}
\fntext[cor1]{fax:+381-11-3160290;tel:+381-11-3713136}
\title{Invisibility in one  dimensional systems:a possible way to achieve it}



\address{Institute of Physics,Pregrevica 118,11080 Zemun-Belgrade,Serbia}

\begin{abstract}

The aim of this paper is to provide the mathematical framework for analyzing the possibility of rendering one dimensional objects visible with difficulty and  ultimately invisible by changing the experimental conditions to which they are subdued. The calculation will be performed using the Hubbard model ,and it will be based on a recent expression for the reflectivity of a particular class of Q1D organic salts.Two sets of material parameters for which the reflectivity of a $1D$ object tends to zero are determined.

\end{abstract}

\end{frontmatter}


\section{Introduction}
\label{}
Rendering real objects and even beings invisible has for many years been in the realm of science fiction. Any kind of device supposed to render an object invisible should lead light around it as if the object was not present. In recent years invisibility became a subject of real studies,in which the usual approach is cloaking objects and thus rendering them invisible (examples are [1],[2]) . For a recent comparison between various cloaking methods see for example [3].

In this paper,a slightly different approach will be taken. In the real world,we see non-radiating objects due to the fact that they reflect incoming light. If an object does not reflect incoming light,it can not be seen. Starting from the expression for the reflectivity of one dimensional systems, derived within the Hubbard model [4] , it will be attempted to determine the experimental conditions for which the reflectivity of a one dimensional object becomes zero or approaches it. The calculation [4] was performed with the aim of applying it to a particular class of organic materials known as Bechgaard salts, which are quasi one dimensional (Q1D). The general chemical formula of these materials is $(TMTSF)_{2}X$ , where $(TMTSF)_{2}$ stands for bi-tetra-methyl-tio-seleno-fulvalene, and $X$ is any anion. Existing results for the real ($\sigma_{R}$)  and immaginary ($\sigma_{I}$) components of the electrical conductivity  of these materials obtained in [5] will be used in the present work.  

This paper consists of several sections. The next one contains a brief outline of the calculation reported in detail in [4]. The third section is devoted to the determination of the derivative of the reflectivity with respect to the temperature,as this is the experimentally most easily controllable parameter. Prelminary results of the calculation are presented in section 4, and the conclusions are briefly stated in section 5.

\section{Theory}
\label{}
Using standard equations of optics, it was shown in [4] that the reflectivity $R$ of a one dimensional ($1D$) system  is approximately given by 
\begin{equation}
	R\cong1-\frac{2}{\pi}\frac{\omega}{K \sigma_{R}}+\frac{1}{2\pi^{3}}\frac{K\omega}{\sigma_{R}^{3}}
\end{equation}
and $K$ is the following function:
\begin{eqnarray}
	K_{{1},{2}}^{2}=\frac{1}{2\omega^{2}}[4\pi\omega\sigma_{I}-\omega^{2}\pm[(4\pi\omega\sigma_{R})^{2}+
	\nonumber\\
	(\omega^{2}-4\pi\omega\sigma_{I})^{2}]^{1/2}]
\end{eqnarray}
The real part of the conductivity is given by the expression (derived in [5]):   
\begin{eqnarray}
\sigma_{R}(\omega_{0})=(1/2\chi_{0})(\omega_{P}^{2}/\pi)[\omega_{0}^{2}-(bt)^{2}]^{-1}\times
\nonumber\\
(Ut/N^{2})^{2}\times S,
\end{eqnarray}
in which the symbol $S$ denotes the function:
\begin{eqnarray}
S=42.49916\times(1+\exp(\beta(-\mu-2t)))^{-2}+78.2557\times
\nonumber\\
(1+\exp(\beta(-\mu+2t\cos(1+\pi))))^{-2}+
\nonumber\\
(bt/(\omega_{0}+bt))\times
(4.53316\times(1+\exp(\beta(-\mu-2t)))^{-2}+
\nonumber\\
24.6448(1+\exp(\beta(-\mu+2t\cos(1+\pi)))))^{-2})
\nonumber\\
\end{eqnarray}
The symbol $\mu$ denotes the chemical potential of the electron gas, determined in [5]: 
\begin{equation}
\mu=\frac{(ns-1)t(\beta t)^{6}}{1.1029+.1694(\beta t)^{2}+.0654(\beta t)^{4}}	
\end{equation}
where $b=- 4(1+\cos(1-\pi))$, $s$ the lattice constant, $\omega_{0}$ the real part of the frequency, $n$ is the mean number of electrons per lattice site, $\beta$ the inverse temperature and 
all the other symbols have meanings standard within the Hubbard model. The immaginary part of the conductivity is given by [5]:
\begin{equation}
	\sigma_{I}=\frac{\omega_{P}^{2}}{4\pi\omega} [1-\frac{\chi_{R}}{\chi_{0}}]
\end{equation}
where 
\begin{equation}
	\chi_{R}\cong\frac{128U^{2}t^{3}\cos^{2}\frac{1-2\pi}{2}}{(1+\exp(\beta(-\mu-2t))^{2})(\omega+2bt)N^{4}}+...
\end{equation}

\section{The calculations}
\label{}
The experimental parameter which is most easily controlled is the temperature of the specimen. In order to theoretically explore the influence of variable temperature on the reflectivity $R$, it is useful to represent the reflectivity as

\begin{equation}
R(T)\cong R(T_{0}) + \frac{\partial R}{\partial T} (T-T_{0}) 
\end{equation}
Obviously 
\begin{equation}
	\frac{\partial R}{\partial T}= \frac{\partial R}{\partial \sigma_{R}} \frac{\partial \sigma_{R}}{\partial T}= \frac{\partial R}{\partial \sigma_{R}} \frac{\partial \sigma_{R}}{\partial \beta}\frac{\partial \beta}{\partial T} 
\end{equation}
It follows from eq.(1) that
\begin{eqnarray}
\frac{\partial R}{\partial \sigma_{R}}= \frac{K}{2} (\frac{\partial \sigma_{R}}{\partial \omega})^{-1} \frac{1}{(\pi\sigma_{R})^{3}}[1-(\frac{2\pi\sigma_{R}}{K})^{2}]+\frac{\omega}{2} \frac{1}{(\pi\sigma_{R})^{3}}\frac{\partial K}{\partial\sigma_{R}}\nonumber\\\times[1+(\frac{2\pi\sigma_{R}}{K})^{2}]- \frac{3K}{2}\frac{\omega}{\pi^{3}\sigma_{R}^{4}}\times[1-\frac{1}{3}(\frac{2\pi\sigma_{R}}{K})^{2}] 
\end{eqnarray}
It can be shown that
\begin{eqnarray}
\frac{\partial K}{\partial\sigma_{R}}\cong 4\sqrt{2}\pi^{2}\sigma_{R}[(4\pi\sigma_{R}\omega)^{2}+(\omega^{2}-4\pi\sigma_{I}\omega)^{2}]^{-1/2}\nonumber\\
\times[\frac{4\pi\sigma_{I}\omega-\omega^{2}+\sqrt{(4\pi\sigma_{R}\omega)^{2}+(\omega^{2}-4\pi\sigma_{I}\omega)^{2}}}{\omega^{2}}]^{-1/2}
\end{eqnarray}
and from eqs.(3),(4) it follows that:
\begin{eqnarray}
\frac{\partial \sigma_{R}}{\partial \beta} \cong \frac{U^{2}\omega_{P}^{2} t^{3}}{N^{4}\chi_{0}(\omega^{2}-b^{2}t^{2})}[27.0558\frac{\exp[\beta(-2t-\mu)]}{(1+\exp[\beta(-\mu-2t)])^{3}}+\nonumber\\ 26.9174\frac{\exp[\beta(-2t\cos(1)-\mu)]}{(1+\exp[\beta(-\mu-2t\cos(1))])^{3}}]+\ldots
\end{eqnarray}

and also
\begin{eqnarray}
	\frac{\partial \sigma_{R}}{\partial \omega}\cong \frac{- 13.5279 t^{2}U^{2}\omega\omega_{P}^{2}}{\chi_{0}N^{4}(\omega^{2}-(b t)^{2})^{2}[1+\exp[\beta(-2t-\mu)]]^{2}}-\nonumber\\
\frac{24.9096t^{2}U^{2}\omega\omega_{P}^{2}}{\chi_{0}N^{4}(\omega^{2}-(b t)^{2})^{2}[1+\exp[\beta(-2t\cos(1)-\mu)]]^{2}}+...
\end{eqnarray}
\section{Preliminary results}
\label{} 
Using expressions given in the preceeding section it would be possible in principle to obtain an analytical expression for the derivative $\frac{\partial R}{\partial T}$. The full expression for this derivative which one would get in this way would be too long for any meaningful analysis, because it would have more than 250 terms. The first of these terms is:
 
\begin{eqnarray}
	\frac{\partial R}{\partial T}\cong-282562\frac{t^{3}\omega(U\omega_{P})^{2}\exp(\beta(-2t-\mu))}{[(1+\exp(\beta(-2t-\mu)))^{3}N^{4}T^{2}\chi_{0}\sigma_{R}^{2}(\omega^{2}-(bt)^{2})]}\nonumber\\
\times[(4\pi\sigma_{R}\omega)^{2}+(\omega^{2}-4\pi\sigma_{I}\omega)^{2}]^{-1/2}\nonumber\\
\times[\frac{4\pi\sigma_{I}-\omega}{\omega}+\frac{[(4\pi\sigma_{R}\omega)^{2}+(\omega^{2}-4\pi\sigma_{I}\omega)^{2}]^{1/2}}{\omega^{2}}]^{-1/2}
\end{eqnarray}
Various terms in this expressions have their meanings standard within the Hubbard model. The real $\sigma_{R}$ and immaginary $\sigma_{I}$ parts of the conductivity, as well as the chemical potential $\mu$ can be calculated by equations $(3)$ to $(7)$,details of which are discussed in [5]. 

However,conclusions can be drawn by analyzing graphically the full expression for the derivative of $R$ with respect to $T$.The values of the material parameters can be chosen freely. The same values were chosen as in [5],that is: $U=4t$,$\omega_{P}=3t$,$N=150$,$\chi_{0}=1/3$,$s=1$,$\omega=0.04$,$t=0.015$ and $n$ was left as a free parameter. Note that in the usual terminology of the Hubbard model, $n=1$ corresponds to half filling. From the experimental point of view, $n=1$ corresponds to a pure specimen,while values different from $1$ correspond to doping by electron donors or acceptors. 
Calculating the value of the derivative $\frac{\partial R}{\partial T}$, one can get curves of various shapes. The following three figures show the temperature dependence of the derivative $\frac{\partial R}{\partial T}$ for three values of the mean number of electrons per site: ($n=0.75$ on fig.1), ($n=0.95$ on fig.2) and ($n=1.1$ and $n=1.15$ on fig.3). The curves are normalized to $\frac{\partial R}{\partial T}=1$ at $T=100K$. 

Interesting conclusions can be drawn by mutually comparing figures 1 to 3. The first two figures correspond to $n<1$. The shapes of the curves of figs.1 and 2 are clearly similar. Note that on both of these figures there exists a "`transition point"' -a value of the temperature at which the derivative $\frac{\partial R}{\partial T}$ changes sign. For $n=0.75$ this occurs for $T\approx160 K-170 K$,while for $n=0.95$ the transition occurs for $T\approx125 K$.There is no such a point on the curves drawn for $n>1$.

If the derivative $\frac{\partial R}{\partial T}$ is negative, and the difference $T-T_{0}$ is positive, the reflectivity  diminishes and at some value of the temperature becomes close to or equal to zero. The fact that $R\rightarrow0$ means that the object becomes visible with increasing difficulty and ultimately invisible.

The third figure was prepared for $n=1.1$ and $n=1.15$ which experimentally corresponds to doping of a specimen with electron donors. For the parameters used in preparing fig.3, the derivative of the reflectivity with respect to the temperature is positive. In order to make the reflectivity diminish,according to eq.(8), the difference $T-T_{0}$ has to be negative. The implication for the reflectivity and invisibility is that in this case the object has to be cooled in order to render it visible with difficulty and ultimately invisible. This illustrates an extremely finely tuned dependence of the derivative of $R$ with respect to $T$ on all the parameters of the problem. The temperature at which the reflectivity can be expected to become zero can be calculated as
\begin{equation}
	T=T_{0}-(\frac{\partial R}{\partial T})^{-1} R(T_{0}) 
\end{equation}
which can easily be obtained from eq.(8). 

\begin{center}
\begin{figure}
\includegraphics[width=9cm,height=8cm]{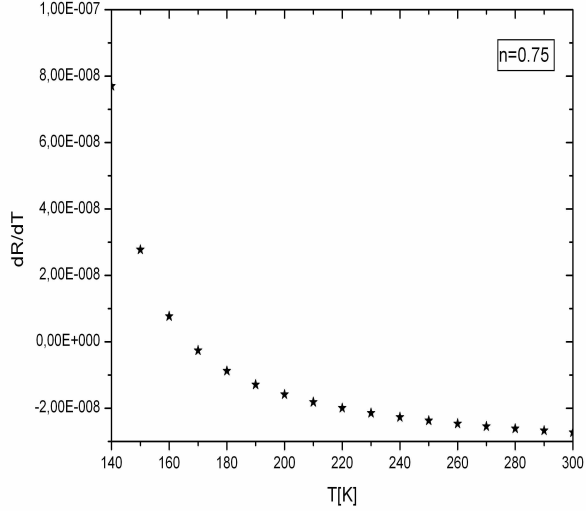}
\caption{The values of $\frac{\partial R}{\partial T}$ in the case of the band filling $n=0.75$}
\end{figure}
\end{center}

\begin{center} 
\begin{figure}
\includegraphics[width=9cm,height=8cm]{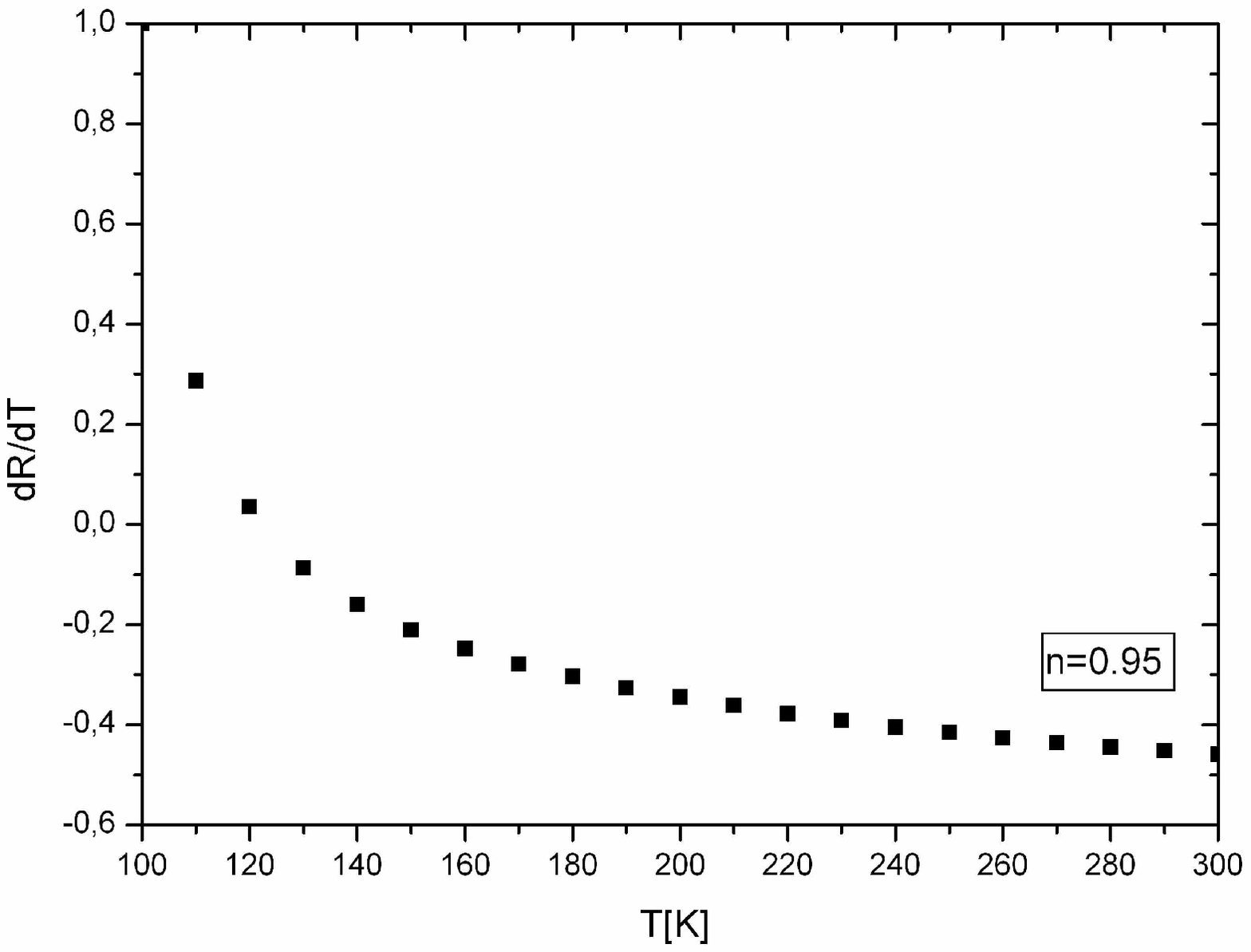}
\caption{The values of $\frac{\partial R}{\partial T}$ in the case of the band filling $n=0.95$}
\end{figure}
\end{center}

\begin{center}
\begin{figure}
\includegraphics[width=9cm,height=8cm]{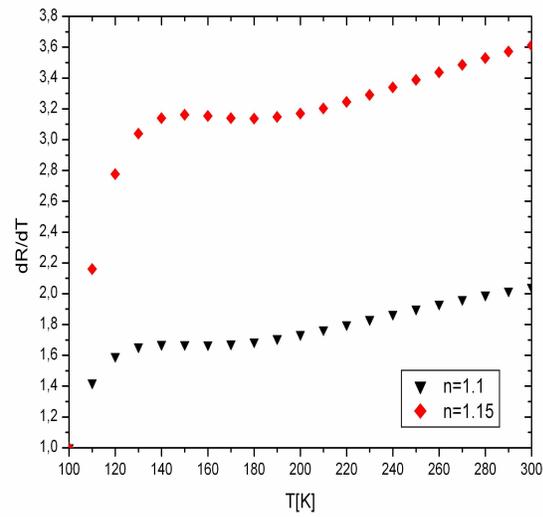}
\caption{The values of $\frac{\partial R}{\partial T}$ in the case of the band filling $n=1.1$ and $n=1.15$}
\end{figure}
\end{center}
\section{Conclusions}
In this paper we have analyzed the problem of rendering 1D objects visible with difficulty or completely invisible. The aim was more to give the necessary mathematical framework than to go into details. The approach taken was based on material science. It was shown that the reflectivity of a 1D object,calculated within the Hubbard model, is finely dependent on the material parameters and the temperature of the specimen.Two sets of material parameters were found for which reflectivity actually goes to 0 if the tempeature of the specimen is changed in a suitable way.

The idea for achieving invisibility discussed in this paper has a distinct advantage over the usual cloaking approach. Namely, when cloaking an object,the obvious question is whether or not the object is visible from behind the cloak. For example,when making a cloaking experiment in two dimensions, the object behind the cloak is visible from the third dimension ([6] and references therein). Such a problem does not appear in the approach discussed here, because invisibility is achieved by choosing the material parameters, which means that the object is invisible from all sides. 

A recent proposal of detecting a perfect invisibility cloak by shooting a particle through it [7],is inapplicable in the case of the method discussed in the present paper. The object is here rendered invisible by the choice of its material parameters,and not by the presence of a cloak.        
\section{Acknowledgement}
\label{}
This paper was prepared within the project No.141007 financed by the Ministry of Science and Technological Developement of Serbia.
{}

\section{References}
\begin{thebibliography}{00}

\bibitem{[1]}
J.Valentine,J.Li,Th.Zentgraf et al,Nature Materials,8 (2009) 568.

\bibitem{[2]}
A.Alu and N.Engheta,J.Opt.A:Pure Appl.Opt,10 (2008) 093002. 

\bibitem{[3]} 
P.Alitalo, H.Kettunen and S.Tretyakov,J.Appl.Phys.,107 (2010) 034905 and

preprint arXiv:0912.3617 v1 (2009).

\bibitem{[4]}
V.Celebonovic,Acta Phys.Polonica  A112 (2007) 949 and 

preprint arXiv 1004.3014.

\bibitem{[5]}
V.Celebonovic in: Trends in Materials Science Research,Editor

B.M.Caruta,Nova Science Publishers Inc.,New York,(2005) pp.245-260.

\bibitem{[6]}
T.Ergin,N.Stenger,P.Brenner et al.,Science,{\bf328},337 (2010). 

\bibitem{[7]} 
B.Zhang and B.-I.Wu.,Phys.Rev.Lett.,{\bf103},243901 (2010). 
  
\end{thebibliography}
\end{document}